\documentclass[11pt,prd,reprint,notitlepage,nofootinbib,superscriptaddress]{revtex4-1}

\usepackage{amsmath,bm}
\usepackage{amsfonts}
\usepackage{graphicx, adjustbox}
\usepackage{aas_macros}
\usepackage{listings}
\usepackage{hyperref}

\usepackage[usenames]{xcolor}
\definecolor{mpl_blue}{HTML}{1F77B4}
\definecolor{mpl_orange}{HTML}{FF7F0E}
\definecolor{mpl_green}{HTML}{2CA02C}
\definecolor{mpl_red}{HTML}{D62728}

\usepackage{orcidlink}

\begin{document}

\title{Efficient gravitational wave searches with pulsar timing arrays using Hamiltonian Monte Carlo}

\author{Gabriel E.~Freedman\orcidlink{0000-0001-7624-4616}}
\affiliation{Center for Gravitation, Cosmology and Astrophysics, University of Wisconsin--Milwaukee, P.O. Box 413, Milwaukee, Wisconsin 53201, USA}

\author{Aaron D.~Johnson\orcidlink{0000-0002-7445-8423}}
\affiliation{Center for Gravitation, Cosmology and Astrophysics, University of Wisconsin--Milwaukee, P.O. Box 413, Milwaukee, Wisconsin 53201, USA}
\affiliation{Theoretical AstroPhysics Including Relativity (TAPIR), MC 350-17, California Institute of Technology, Pasadena, California 91125, USA}

\author{Rutger van Haasteren\orcidlink{0000-0002-6428-2620}}
\affiliation{Max-Planck-Institut f\"ur Gravitationsphysik (Albert-Einstein-Institut), Callinstrasse 38, D-30167, Hannover, Germany}

\author{Sarah J.~Vigeland\orcidlink{0000-0003-4700-9072}}
\affiliation{Center for Gravitation, Cosmology and Astrophysics, University of Wisconsin--Milwaukee, P.O. Box 413, Milwaukee, Wisconsin 53201, USA}

\date{\today}

\begin{abstract}
Pulsar timing arrays (PTAs) detect low-frequency gravitational waves (GWs) by looking for correlated deviations in pulse arrival times. Current Bayesian searches use Markov chain Monte Carlo (MCMC) methods, which struggle to sample the large number of parameters needed to model the PTA and GW signals. As the data span and number of pulsars increase, this problem will only worsen. An alternative Monte Carlo sampling method, Hamiltonian Monte Carlo (HMC), utilizes Hamiltonian dynamics to produce sample proposals informed by first-order gradients of the model likelihood. This in turn allows it to converge faster to high dimensional distributions. We implement HMC as an alternative sampling method in our search for an isotropic stochastic GW background, and show that this method produces equivalent statistical results to similar analyses run with standard MCMC techniques, while requiring 100--200 times fewer samples. We show that the speed of HMC sample generation scales as $\mathcal{O}(N_\mathrm{psr}^{5/4})$ where $N_\mathrm{psr}$ is the number of pulsars, compared to $\mathcal{O}(N_\mathrm{psr}^2)$ for MCMC methods. These factors offset the increased time required to generate a sample using HMC, demonstrating the value of adopting HMC techniques for PTAs.
\end{abstract}

\maketitle

\section{Introduction}

Pulsar timing arrays (PTAs) \cite{1978SvA....22...36S, 1979ApJ...234.1100D, 1990ApJ...361..300F} seek to detect low-frequency gravitational waves (GWs) by looking for spatial correlations induced in the times of arrival (TOAs) pulses from millisecond pulsars. PTAs are most sensitive in the nanohertz frequency regime ($\sim$1--100 nHz), where the dominant source of GWs is expected to be a stochastic gravitational wave background (GWB) originating from a cosmic population of supermassive black hole binaries (SMBHBs) \cite{shm+04,2005ApJ...623...23S,rsg2015,stc+19}. The North American Nanohertz Observatory for Gravitational Waves (NANOGrav) \cite{2019BAAS...51g.195R} has been collecting pulsar TOA data since 2004. NANOGrav, along with the European Pulsar Timing Array \cite{2016MNRAS.458.3341D}, Parkes Pulsar Timing Array \cite{2020PASA...37...20K}, and the Indian Pulsar Timing Array Project \cite{2022arXiv220609289T} form the International Pulsar Timing Array (IPTA) \cite{2019MNRAS.490.4666P}.

Detection of low-frequency GWs provides a valuable tool for studying parts of the dynamical universe not accessible through electromagnetic observations. Constraining the GWB shape and strength can provide useful constraints on properties of the SMBHB population including the black hole--host galaxy scaling relations \cite{2015MNRAS.447.2772R, 2013MNRAS.433L...1S} and the astrophysical environments of SMBHBs emitting GWs \cite{1996NewA....1...35Q,Sesana_2006,2009ApJ...700.1952H,10.1111/j.1365-2966.2010.17782.x,scm2015}. The GWB could also contain contributions from more speculative sources such as primordial GWs from inflation \cite{1976JETPL..23..293G, 2016PhRvX...6a1035L} and networks of cosmic strings \cite{2007PhRvL..98k1101S, 2014PhRvD..89b3512B}.

GW signals can be extracted as a correlated signal from pulsar timing data only after subtracting the pulsar's timing model and accounting for underlying sources of noise in both the pulsar and observing instruments. These analyses are frequently done using Bayesian techniques \cite{van_Haasteren_2009,vanHaasteren:2012hj,Lentati:2012xb,2016ApJ...821...13A}, which we outline in Sec. \ref{sec:methods}. In order to perform the Bayesian searches, NANOGrav makes use of the parallel-tempering Markov chain Monte Carlo (MCMC) code \texttt{PTMCMCSampler} \cite{justin_ellis_2017_1037579}, which includes a variety of jump proposal schemes such as differential evolution, prior draws, and adaptive Metropolis.

MCMC methods work adequately for a large portion of statistical models, but simple MCMC algorithms such as random-walk Metropolis \cite{1953JChPh..21.1087M} or Gibbs sampling \cite{Geman&Geman} become slow as the size and complexity of the model grow and take considerably longer to converge. Both of the aforementioned methods use random-walk proposals to generate samples and explore the parameter space, which tend to be increasingly inefficient when the target distribution includes correlations among the parameters \cite{2011hmcm.book..113N}. Hamiltonian Monte Carlo (HMC) \cite{1987PhLB..195..216D, 2011hmcm.book..113N} removes the requirement to sample the model randomly, and replaces it with a simulation of Hamiltonian dynamics on the distribution itself. This scheme allows samples to be drawn at much further distances from one another, and explores the full parameter space in a more efficient way. For a target distribution of dimension $d$, the cost of drawing an independent sample with HMC goes roughly as $\mathcal{O}(d^{5/4})$, compared to $\mathcal{O}(d^{2})$ for random-walk Metropolis \cite{PhysRevD.38.1228}. The no-u-turn sampler (NUTS) \cite{2011arXiv1111.4246H} algorithm provides a basis for performing analysis with HMC without pretuning the sampling.

The HMC algorithm was initially developed for the problem of performing lattice field theory simulations of quantum chromodynamics \cite{1987PhLB..195..216D}. The earliest approach applying HMC to PTA science was in the development of a model-independent method for performing Bayesian analyses on pulsar timing data \cite{2013PhRvD..87j4021L}. The technique worked extremely well when applied to the IPTA Mock Data Challenge.\footnote{The first IPTA Mock Data Challenge was developed by Fredrick Jenet, Kejia Lee, and Michael Keith and administered in 2012.} When applied to real data, however, the sampling could not fully explore the hierarchical model and became stuck in ``Neal's funnel'' \cite{Neal2003SliceS}. Applying data-aware coordinate transformations using the Cholesky decomposition helped deal with hierarchical funneling, and consequently there was a successful application of HMC to the targeted problem of outlier excision from PTA datasets \cite{2017MNRAS.466.4954V}. The trade-off was that the additional transformations made sampling the hierarchical likelihood slower than the typical marginalized likelihood that was already used. As a result, HMC was not further explored in this context and has since remained largely underutilized towards the broad array of PTA science.

In this paper, we present a method for performing PTA GW searches using HMC as the underlying sampling algorithm. This represents the first attempt at applying HMC to the marginalized PTA likelihood, where we can avoid the funneling that plagues hierarchical models while still leveraging the benefits of HMC in exploring high-dimensional distributions. We test this method on the NANOGrav 11-year dataset \cite{2018ApJS..235...37A}, as well as realistic simulated data with similar red and white noise to the NANOGrav 11-year dataset. We demonstrate that performing a Bayesian GWB search with HMC results in a significant reduction in required sample generation to give equivalent results to current methods.

We also show that the additional gradient calculations necessary for HMC to operate scale roughly the same as the current likelihood evaluation with respect to the number of pulsars in a given dataset. Additionally we demonstrate that when comparing the time to generate independent samples, HMC outperforms traditional MCMC methods for PTA models of varying size in accordance with the expected scaling. This is a necessary consideration as the sizes of PTAs will continue to grow and with that the number of parameters needed to sample over.

This paper is organized as follows. In Sec.~\ref{sec:methods}, we describe the methods, signal models, and software used. In Sec.~\ref{sec:results} we present the results of a GWB search using HMC, and compare the accuracy and efficiency of this method for both real and simulated PTA data. We conclude in Sec.~\ref{sec:conclusions} and discuss how this method could be utilized for future PTA work.

\section{Methodology and Software}
\label{sec:methods}

In this section, we provide a brief outline of a typical PTA Bayesian GW search. We then give an overview of the HMC and NUTS algorithms, and discuss how to apply these methods to existing PTA work.

\subsection{PTA signal model}
\label{sec:gwb}

We now discuss the PTA likelihood function. Following the outline provided in \cite{2016ApJ...821...13A}, we start by considering a single pulsar and its timing residual vector $\delta\mathbf{t}$ with length equal to the number of TOAs in our dataset, $N_{\mathrm{TOA}}$. This timing residual data can be decomposed into individual components:
\begin{equation}
    \delta\mathbf{t} = M\boldsymbol{\epsilon} + F\mathbf{a} + U\mathbf{j} + \mathbf{n}.
\end{equation}
Each term describes a different inaccuracy or source of noise that contributes to the residual data. The term $M\boldsymbol{\epsilon}$ represents inaccuracies stemming from the subtraction of the pulsar's timing model, with $M$ the timing model design matrix, and $\boldsymbol{\epsilon}$ the vector of timing model parameter offsets. The effects due to low-frequency (``red'') noise are encoded in the term $F\mathbf{a}$. We choose to define this in a rank-reduced basis where $F$ represents our matrix of basis functions, in this case alternating sine and cosine functions, and $\mathbf{a}$ represents a set of Fourier coefficients. The term $U\mathbf{j}$ describes noise that is completely uncorrelated in time but completely correlated across observations of a similar epoch. The matrix $U$ maps between $N_{\mathrm{TOA}}$ residual data and $N_{\mathrm{epoch}}$ observation sessions, and $\mathbf{j}$ accounts for the correlated noise in each epoch. The final term, $\mathbf{n}$, includes any other high-frequency (``white'') noise that cannot be accounted for in the previous terms, such as radiometer noise.

Previous Bayesian analysis schemes \cite{2009MNRAS.395.1005V, 2010MNRAS.401.2372V, 10.1111/j.1365-2966.2011.18613.x, 2013CQGra..30v4004E, 2013ApJ...769...63E} have described the white noise with EFAC (constant multiplier to TOA uncertainties) and EQUAD (white noise added in quadrature to EFAC) parameters and employed a power-law model to describe the red noise. The sum of these white noise covariances we describe via a matrix $N$. The parameters describing $\boldsymbol{\epsilon}$, $\mathbf{a}$, and $\mathbf{j}$ we group as follows:
\begin{equation}
    T = \begin{bmatrix}
    M & F & U
    \end{bmatrix}, \quad \mathbf{b} = \begin{bmatrix}
    \boldsymbol{\epsilon} \\ \mathbf{a} \\ \mathbf{j}
    \end{bmatrix}.
\end{equation}
We place a Gaussian prior on these parameters with covariance:
\begin{equation}
    B = \begin{bmatrix}
    \infty & 0 & 0 \\
    0 & \varphi & 0 \\
    0 & 0 & \mathcal{J}
    \end{bmatrix},
\end{equation}
where $\infty$ represents a diagonal matrix of infinities corresponding to unconstrained uniform priors on all timing model parameters. The parameters that describe $\mathcal{J}$ we refer to as ECORR and correspond to the epoch-correlated white noise signals per receiving back end. The matrix $\varphi$ defines the parameters involving red noise signals, which includes low-frequency noise intrinsic to each pulsar, as well as the stochastic GWB. For this paper, we performed our analysis by modeling the GWB using a fiducial power-law spectrum of the characteristic GW strain $h_{c}$ and cross-power spectral density $S_{ab}$:
\begin{align}
    h_{c}(f) &= A_{\mathrm{gw}}\left(\frac{f}{f_{yr}}\right)^{\alpha}, \\
    S_{ab}(f) &= \Gamma_{ab}\frac{A_{\mathrm{gw}}^{2}}{12\pi^{2}} \left(\frac{f}{f_{\mathrm{yr}}}\right)^{-\gamma} f_{\mathrm{yr}}^{-3},
\end{align}
where $\gamma = 3 - 2\alpha$. For a background generated by the GW emission from the evolution of a population of inspiraling SMBHBs in circular orbits, we have $\alpha=-2/3$, which implies $\gamma=13/3$ \cite{2001astro.ph..8028P}. The function $\Gamma_{ab}$ is called the overlap reduction function (ORF) and describes the average correlations between any two pulsars $a$ and $b$ as a function of their angular separation. For an isotropic, stochastic GWB, this ORF is given by 
the Hellings-Downs correlation: \cite{1983ApJ...265L..39H} %\cite{1983grg1.conf..963H}

\begin{equation}\label{eq:hd}
    \Gamma_{ab} = \frac{3}{2}x_{ab}\ln x_{ab} - \frac{x_{ab}}{4} + \frac{1}{2} + \frac{\delta_{ab}}{2} \,,
\end{equation}
where $x_{ab} = \left(1 - \cos \xi_{ab}\right) / 2$ for two pulsars with angular separation $\xi_{ab}$.

We analytically marginalize over the timing model parameters to reduce the overall dimensionality of our posterior \cite{2013PhRvD..87j4021L, 2014PhRvD..90j4012V} and are left with the form of the likelihood that is used for the analysis in this paper:
\begin{equation}\label{eq:likelihood}
    \mathcal{P}(\delta\mathbf{t} | \phi) =  \frac{\exp\left(-\frac{1}{2}\delta\mathbf{t}^{T}C^{-1}\delta\mathbf{t}\right)}{\sqrt{\det 2\pi C}},
\end{equation}
where $C = N + TBT^{T}$. We define $\phi$ as the set of all varying parameters in our model. We compute the likelihood and perform Bayesian searches using the NANOGrav package \texttt{enterprise} \cite{enterprise}.

\subsection{Hamiltonian Monte Carlo}
\label{sec:hmc}

We now provide a description of the HMC algorithm. In HMC \cite{1987PhLB..195..216D, 2011hmcm.book..113N}, we start by introducing an auxiliary momentum variable $p_{i}$ alongside each target parameter $q_{i}$. In most implementations, the momenta are chosen to be independent of the $q_{i}$ and follow a zero-mean Gaussian distribution, with a covariance matrix $M$ that is typically taken to be the identity. The log of the joint density of $\mathbf{p}$ and $\mathbf{q}$ defines our Hamiltonian:
\begin{equation}
    H\left(\mathbf{p}, \mathbf{q}\right) = U\left(\mathbf{q}\right) + K\left(\mathbf{p}\right) = -\mathcal{L}(\mathbf{q}) + \frac{1}{2}\mathbf{p}^{T}M^{-1}\mathbf{p},
\end{equation}
where $\mathcal{L}(\mathbf{q}) \equiv \log \mathcal{P}(\delta\mathbf{t} | \phi)$ is the log of the likelihood function for the distribution of our target parameters $\mathbf{q}$. Analogous to Hamiltonian dynamics, we have a potential energy term $U(\mathbf{q})$ and a kinetic energy term $K(\mathbf{p})$. We then simulate the evolution of this system over time according to Hamilton's equations:

\begin{equation}
    \frac{dq_{i}}{dt} = \frac{\partial H}{\partial p_{i}}, \qquad
    \frac{dp_{i}}{dt} = -\frac{\partial H}{\partial q_{i}}.
\end{equation}

This can be solved numerically using a symplectic integrator such as a ``leapfrop'' method, which for an integration step size $\varepsilon$ uses an update scheme:
\begin{subequations}\label{eq:leapfrog}
\begin{align}
    \mathbf{p}^{t + \varepsilon/2} &= \mathbf{p}^{t} + \left(\frac{\varepsilon}{2}\right)\nabla_{\mathbf{q}}\mathcal{L}(\mathbf{q}^{t}),
    \\
    \mathbf{q}^{t+\varepsilon} &= \mathbf{q}^{t} + \varepsilon\mathbf{p}^{t+\varepsilon/2},
    \\
    \mathbf{p}^{t + \varepsilon} &= \mathbf{p}^{t+\varepsilon/2} + \left(\frac{\varepsilon}{2}\right)\nabla_{\mathbf{q}}\mathcal{L}(\mathbf{q}^{t+\varepsilon}),
\end{align}
\end{subequations}
where superscripts denote the time at which the particular quantity is evaluated. The standard method for producing a chain of samples using HMC then proceeds as follows: We first resample our momenta distribution. Then for a set number of leapfrog steps $L$, we use Eq. \eqref{eq:leapfrog} to evolve our system through time and propose some final position and momentum vectors $\tilde{\mathbf{q}}$ and $\tilde{\mathbf{p}}$. This proposal is accepted or rejected according to the Metropolis algorithm \cite{1953JChPh..21.1087M}.

Mapping the path of the leapfrog integrator leads to a useful sanity check of HMC: trajectory divergences. These divergences occur when the trajectory taken via Hamiltonian simulation departs from the true trajectory, and risk biasing estimates or reducing HMC to random-walk behavior \cite{2016arXiv160400695B}. By tracking the trajectories and alerting the user of large divergences, HMC offers another diagnostic to detect unsuitably parametrized models that is not possible with Metropolis-Hastings (MH) MCMC methods.

There are limitations to HMC and the models under which it can be used properly. Due to its origins in Hamiltonian dynamics, HMC can only operate in continuous state spaces and contains no internal recourse to deal with discrete variables. In such cases, the discrete variables can be handled with separate algorithms such as Gibbs sampling \cite{Geman&Geman}. HMC also requires that the log density of the target distribution is differentiable almost everywhere with respect to the model parameters, with the exception coming at points of probability 0 \cite{2011hmcm.book..113N}. Additionally, HMC struggles when there is strong multimodality in the target distribution due to the modes being separated by regions of very low probability \cite{2011PatRe..44.2738S}. The PTA models used in this paper satisfy the above conditions, and HMC remains a valid choice of underlying sampling algorithm.

\subsection{No-u-turn sampler}
\label{sec:nuts}

The performance of the HMC algorithm is particularly sensitive to two user-defined parameters: the number of leapfrog steps $L$ and integration step size $\varepsilon$, defined in the above section. If these parameters are not properly tuned, the algorithm may waste computation time or begin to exhibit unwanted random walk behavior and in some cases may not even be ergodic \cite{2011hmcm.book..113N}. In general, tuning these parameters appropriately would require multiple preliminary runs.

The no-u-turn samlper \cite[NUTS;][]{2011arXiv1111.4246H} offers an extension to the HMC algorithm that dynamically tunes the number of leapfrog steps $L$. NUTS uses a recursive doubling algorithm, similar to the one outlined in \cite{Neal2003SliceS}, to determine when the generated proposal trajectory begins to double back on itself, or make a ``U turn''. The algorithm builds a binary tree, simulating Hamiltonian dynamics forwards and backwards randomly in time for $2^{j}$ steps, with $j$ the height of the full tree. If we define $\mathbf{q}^{+}$, $\mathbf{p}^{+}$ and $\mathbf{q}^{-}$, $\mathbf{p}^{-}$ as the position-momenta pairs of the left- and rightmost nodes of the bottom subtree, then the stopping condition for NUTS can be written as:

\begin{equation}
    \left(\mathbf{q}^{+} - \mathbf{q}^{-}\right) \cdot \mathbf{p}^{-} < 0
    \quad\text{or}\quad
    \left(\mathbf{q}^{+} - \mathbf{q}^{-}\right) \cdot \mathbf{p}^{+} < 0.
\end{equation}

The above procedure adaptively tunes the parameter $L$ for each iteration in the chain. The step size parameter $\varepsilon$ in NUTS is set using the method of stochastic optimization with varying adaptation \cite{Andrieu2008ATO}. In particular, Hoffman and Gelman utilize the primal-dual averaging algorithm proposed by \cite{Nesterov2009PrimaldualSM}. With $L$ and $\varepsilon$ automatically tuned, NUTS can be run without any human intervention.

\subsection{Coordinate transformations and software}
\label{sec:software}

Previous approaches to pulsar timing analyses with HMC utilized a hierarchical PTA likelihood. Initially these methods did not include coordinate transformations on the data, and as a result became stuck with hierarchical funneling. This funneling originates from the fact that within hierarchical models, random variables are very highly correlated when the data are sparse \cite{2013arXiv1312.0906B}. One can reduce the correlations between the random variables, and hence the funneling, by adopting a noncentered reparametrization of the data \cite{noncentral_param}. In regards to the hierarchical PTA likelihood, such a reparametrization using the Cholesky decomposition allowed HMC sampling to proceed but at the cost of slowing down the likelihood.

In this paper we are focused entirely on the marginalized PTA likelihood and can therefore leave behind the coordinate transformations designed for hierarchical models. We do employ a set of transformations designed to improve the performance of the NUTS algorithm. First we perform an interval transform, moving all parameters with bounded priors from their interval $[a,b]$ to the whole real line. We then whiten the data using Cholesky whitening to move to a set of transformed variables whose covariance matrix is the identity. This is accomplished through the Hessian calculated around the maximum \textit{a posteriori} parameter vector. Neither transformation considerably alters the likelihood computation speed.

When determining the speed and efficiency of the HMC and NUTS pipeline, one must depend almost entirely on the ability to calculate gradients of the likelihood and do so as quickly as possible. Numerical derivatives are comparably easier to write but slow in practice and prone to errors from approximations. By-hand analytic derivatives are fast but difficult to write into concise code for all but the simplest of models. An excellent solution for arbitrary likelihood functions and their gradients is the package \texttt{JAX} \cite{jax2018github}, which leverages both automatic differentiation and just-in-time compilation to efficiently differentiate native Python code and turn an otherwise slow gradient function into incredibly fast machine executables.
The use of this technique is rather new, with \texttt{JAX} only recently becoming a mature code base, and consequently this marks the first time \texttt{JAX} and automatic differentiation have been utilized for HMC sampling of PTA data.

Summarizing the software used for the analyses to follow in this paper, the signal models and likelihood used in our analyses come from NANOGrav's flagship PTA analysis suite \texttt{enterprise} \cite{enterprise}. We utilize the automatic differentiation capabilities in \texttt{JAX} \cite{jax2018github} to calculate the likelihood derivatives required for HMC to operate. We perform two coordinate transformations on our data to better interface with the NUTS algorithm. Lastly, for the sampling we use a custom-built NUTS code that is freely and openly available in \texttt{piccard}.\footnote{\href{https://github.com/vhaasteren/piccard}{https://github.com/vhaasteren/piccard}} The combination of these three codes leads to an end-to-end pipeline for performing PTA analyses with HMC sampling.

\section{Results}
\label{sec:results}

In this section, we study the HMC sampling method both in its ability to accurately perform Bayesian searches for a stochastic GWB using PTA data, as well the efficiency of such a method when compared against the existing techniques employed by NANOGrav.

The GWB model that is analyzed in this paper arises from a PTA consisting of data from 45 pulsars. The parameters encompassing the signal model closely mimic those outlined in Sec.~\ref{sec:gwb}. We fix white noise parameters to their maximum likelihood values as obtained from individual pulsar noise runs. We model pulsar-intrinsic red noise with a power-law power spectral density (PSD) containing two search parameters $\log_{10}\, A_\mathrm{red} \in U[-18, -11]$ and $\gamma_{\mathrm{red}} \in U[0, 7]$. We model the GWB as a power-law PSD process that is common amongst all the pulsars. The corresponding parameters are an amplitude with log-uniform prior $A_\mathrm{CP} \in U[-18, -12]$ and a spectral index $\gamma_{\mathrm{CP}}$ that we fix to $13/3$. We do not include spatial correlations in our GWB model. This results in a total of $2N_{\text{psr}} + 1$ varying parameters in the model.

We also generate a set of simulated PTA datasets using \texttt{libstempo} \cite{libstempo}. We inject both per-pulsar white and red noise parameters at their maximum likelihood values. The injected values again originate from individual pulsar noise runs, where all parameters for a given pulsar are allowed to vary. Again we include a common process signal representing the GWB with both a fixed amplitude and spectral index at $\log_{10}\, A_\mathrm{CP} = -14.7$ and $\gamma_{\mathrm{CP}} = 13/3$, and do not include interpulsar spatial correlations. We repeat the above procedure for 100 realizations of the GWB which results in a collection of 100 realistic simulated PTA datasets. When analyzing the simulated data, we use a similar signal model to the one described above but this time allow the common-process spectral index to vary as $\gamma_{\mathrm{CP}} \in U[0, 7]$.

Runs conducted with the MH MCMC algorithm use the \texttt{PTMCMCSampler} \cite{justin_ellis_2017_1037579} code. The sampler is set up in similar fashion to the NANOGrav 11-year GWB search \cite{2018ApJ...859...47A}. We include adaptive Metropolis and differential evolution jump proposals. For all varying parameters present in the model, we also add prior draw jump proposals. We do not utilize parallel-tempering in this work.

\subsection{NANOGrav 11-year data comparison}
\label{sec:11year}

\begin{figure}
    \includegraphics[width=\columnwidth]{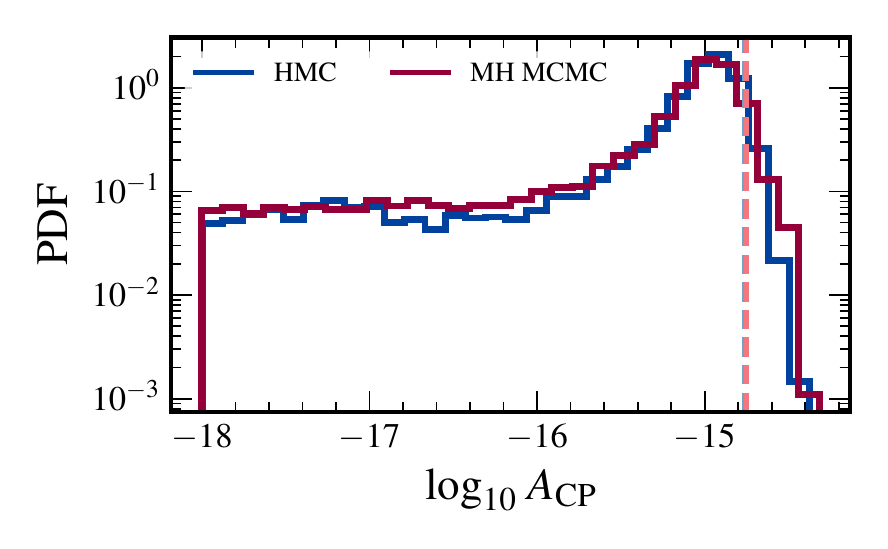}
    \caption{Posterior probability distributions for the amplitude $\log_{10}\, A_{\mathrm{CP}}$ of a common-process signal run using either MH MCMC or HMC as the primary sampling method, computed using the NANOGrav 11-year dataset. The common-process amplitude parameter is set with a log-uniform prior, the common-process spectral index is fixed at 13/3, and no spatial correlations are included. Vertical lines represent 95\% upper limits calculated for posteriors generated using HMC [blue; $A_{\text{CP, HMC}} < 1.72(4) \times 10^{-15}$] and MH MCMC [red; $A_{\text{CP, HMC}} < 1.74(3) \times 10^{-15}$], though the two lines will be difficult to individually resolve due to the similarity in upper limits. We conclude that the two procedures produce consistent posteriors when applied to identical models.}
    \label{fig:corner}
\end{figure}

We perform a stochastic GWB search with both the HMC and MH MCMC algorithms on the NANOGrav 11-year dataset \cite{2018ApJS..235...37A}. This dataset encompasses the timing data for 45 millisecond pulsars. Figure \ref{fig:corner} shows the posterior distributions for the background amplitude $A_{\text{CP}}$ calculated using both Monte Carlo methods. We calculate $95\%$ upper limits on $A_{\text{CP}}$ and estimate uncertainties with bootstrap methods \cite{Efron1979BootstrapMA}. The HMC algorithm produces results that are consistent with the base MH MCMC search, with corresponding $95\%$ upper limits $A_{\text{CP, HMC}} < 1.72(4) \times 10^{-15}$ and $A_{\text{CP, MH MCMC}} < 1.74(3) \times 10^{-15}$.

The MH MCMC sampling routine was run for a total number of samples $M_{\text{MH MCMC}} = 1,000,000$, whereas the HMC routine was run for $M_{\text{HMC}} = 8,000$. The wall time for the MH run was approximately 4 hours, compared to just under 4 hours for the HMC run. Both sets of chains are checked for convergence using the Gelman-Rubin R-hat convergence test \cite{2011arXiv1111.4246H}. It is worth reinforcing that the benefit of generating fewer samples is partially outweighed by the increased computational cost of proposing a new HMC sample. We explore the scaling of sample generation time in Sec.~\ref{sec:speed}.

We also perform a direct comparison to the upper limit calculated in the NANOGrav 11-year GWB search \cite{2018ApJ...859...47A}. In order to do such a comparison, we alter our signal model slightly to match that of the 11-year analysis and adjust the common-process amplitude from a log uniform to a uniform prior $A_{\mathrm{CP}} \in [10^{-18}, 10^{-12}]$. Performing this analysis with the HMC pipeline, again with $M=8,000$ samples, recovers a $95\%$ upper limit of $A_{\text{CP, HMC}} < 1.64(3) \times 10^{-15}$. This is in relative agreement with the result in \cite{2018ApJ...859...47A} of $A_{\text{CP}} < 1.61(2) \times 10^{-15}$ for a similar model with identical Jet Propulsion Laboratory (JPL) ephemeride DE436.

\begin{figure}
    \includegraphics[width=\columnwidth]{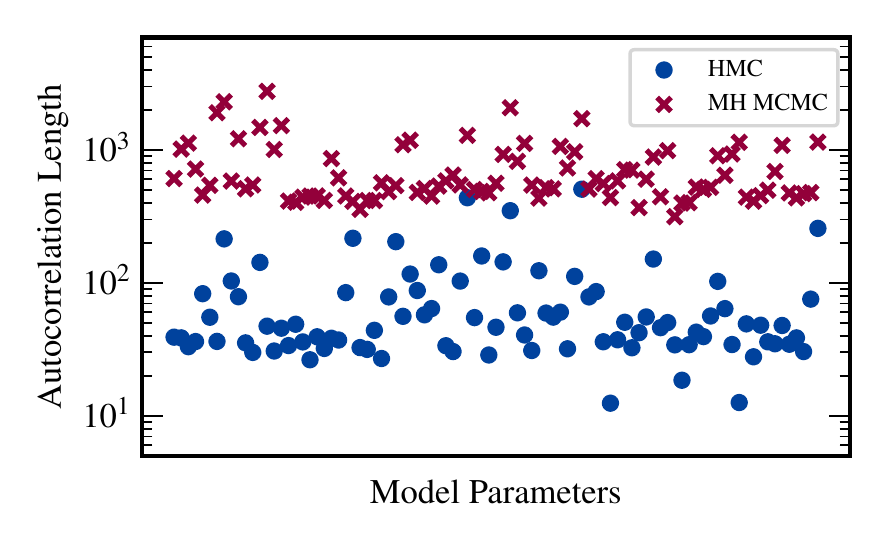}
    \caption{Autocorrelation lengths for 91 parameters ($2N_{\text{psr}}$ individual pulsar red-noise parameters and a common process signal parametrized with an amplitude $A_{\mathrm{CP}}$ and spectral index $\gamma_{\mathrm{CP}}=13/3$) present in a standard GWB model. The autocorrelation lengths are calculated from two sets of chains generated from sampling this model: one sampled with HMC (blue) and one with MH MCMC (red). Each mark represents the approximate number of steps one must jump through that particular parameter's chain to reach an independent sample.}
    \label{fig:autocor}
\end{figure}

We further compare the efficiency of HMC sampling by looking at the autocorrelation lengths of the two sets of chains, measuring how far one must jump through the chain to find the next statistically significant sample. The autocorrelation lengths are calculated per parameter in the model. This was calculated for each set of chains generated with the two Monte Carlo sampling methods, and the results are shown in Fig.~\ref{fig:autocor}. We find that the HMC chains have autocorrelation lengths between 1 and 2 orders of magnitude smaller than those of identical parameters in the MH MCMC chains. This behavior is expected, as the HMC algorithm is designed to take larger, more-informed steps to avoid random walklike behavior and produce a higher ratio of independent samples.

\subsection{Simulated data and parameter recovery}
\label{sec:simulated}

We also aim to test that the HMC algorithm behaves similarly to the standard MH MCMC technique when considering statistical coverage of a standard PTA model. To determine the capability of the sampling methods to accurately recover injected parameters, we consider 100 simulated PTA datasets and seek to verify if in $p$\% of the realizations the injected parameter values fall within the $p$\% credible region of the posteriors. We run standard Bayesian searches on all realizations using both sampling methods.

\begin{figure}
    \includegraphics[width=\columnwidth]{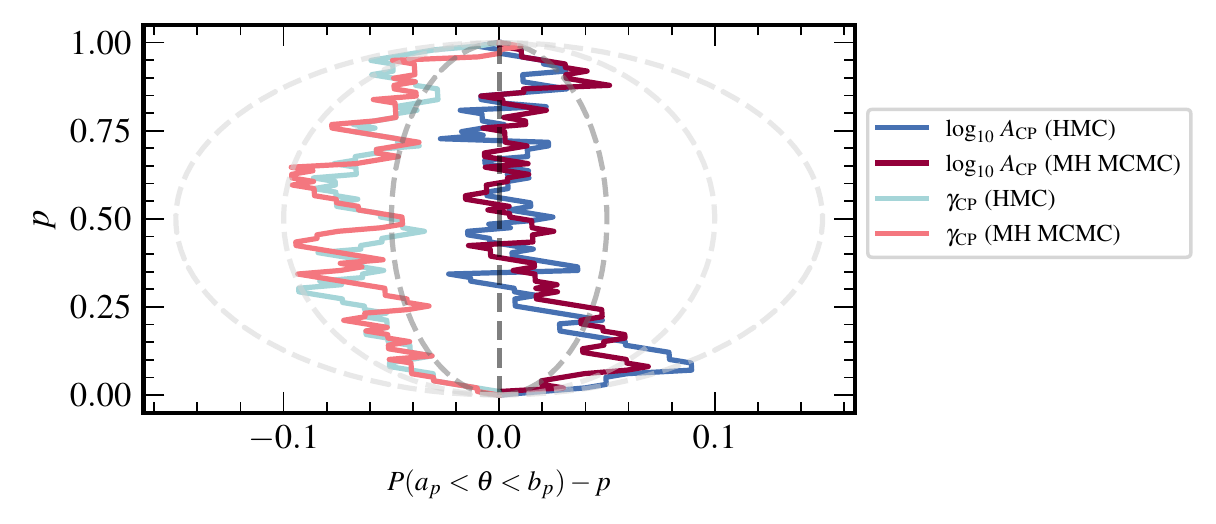}
    \caption{$p-p$ comparison of GWB parameter recovery for both the HMC and MH MCMC sampling methods operating on simulated PTA data. The $x$ axis shows the difference between the fraction of realizations with which the injected values fall within the $p$\% credible region of the posteriors and the $p$\% credible region on the $y$ axis. The vertical dark gray line at $x=0$ represents a perfect recovery of the injected parameter values. The light gray lines represent $1\sigma$, $2\sigma$, and $3\sigma$ deviations.}
    \label{fig:simulations}
\end{figure}

The results of the parameter recovery test described above are summarized in Fig.~\ref{fig:simulations}, with a particular focus on the two parameters describing the GWB. The HMC sampler recovers the injected GWB parameter values with the same consistency as the traditional analysis. Neither method recovers the injected parameters exactly, and therefore no line in Fig.~\ref{fig:simulations} falls directly on the vertical line at $x=0$. This is due to an inherent model mismatch present when simulating data with \texttt{libstempo} and recovering the posteriors separately with \texttt{enterprise}. The simulated GWB is generated with more frequencies than is searched over during the analysis, leading to a natural bias in recovery.\footnote{For further details, see documentation for GWB simulation in the \texttt{toasim} module of \texttt{libstempo}.}

\subsection{Scaling of gradient computation speed}
\label{sec:speed}

\begin{figure}
    \includegraphics[width=\columnwidth]{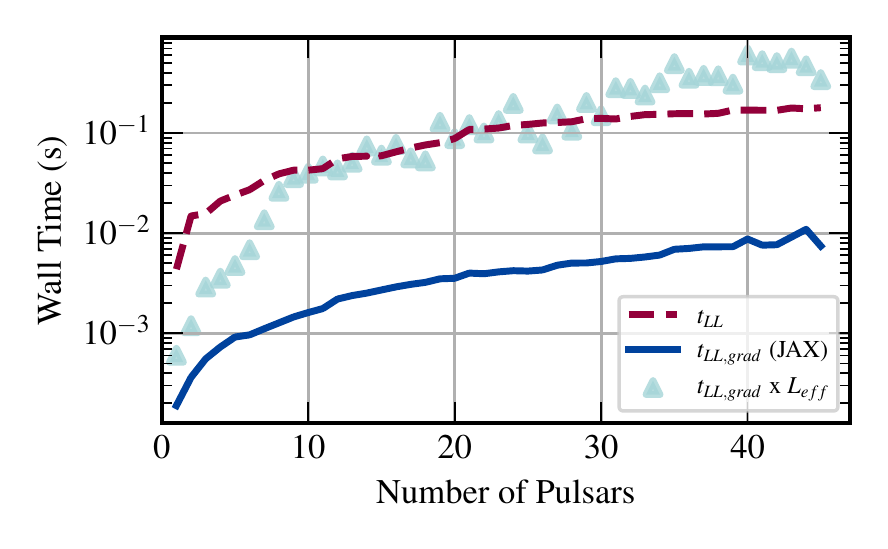}
    \caption{Wall time for calculating implementations of both the log of the PTA likelihood as well as its gradient, scaled by the number of pulsars present in a given model. The red dashed line represents the log-likelihood evaluation as present in the standard PTA analysis suite \texttt{enterprise}. The solid blue line shows the evaluation of the log likelihood and gradient function after being precompiled with \texttt{JAX}. The cyan triangles denote the evaluation times present in the blue line multiplied by a value $L_{\text{eff}}$ representing the effective number of gradient evaluations required to generate a new HMC sample.}
    \label{fig:speed}
\end{figure}

The time per HMC sample generation is dominated by the time to calculate the gradient of the log likelihood necessary for leapfrog integration. The evaluation time for the base likelihood calculation present in $\texttt{enterprise}$ is calculated by averaging the evaluation time for 50 calls of the log likelihood function. We first use a PTA with only a single pulsar, and repeat the above step adding one additional pulsar at a time up to $N_{\text{psr}} = 45$. This produces an idea of how the base likelihood evaluation time, and by extension the MCMC sample generation time, scales with the number of pulsars present in a PTA (Fig.~\ref{fig:speed}: dashed red line).

In order to accurately scale the computation time necessary to draw a sample with NUTS, we must account for the dynamic tuning of the HMC hyperparameter $L$ and note that we likely require multiple evaluations of the log likelihood and gradient to generate a sample. First we consider the evaluation time of the log likelihood and gradient function compiled with \texttt{JAX}, and scale per pulsar following the same procedure defined above (Fig.~\ref{fig:speed}: solid blue line). We then take the 45 separate PTA objects and run standard GWB analyses, with models defined in Sec.~\ref{sec:results}, through the HMC pipeline for $M=10,000$ samples. The height $j$ of the NUTS binary tree defines a total of $L=2^{j}+1$ gradient evaluations per new sample. By averaging this over the full run, we can approximate an $L_{\text{eff}}$ and more accurately scale the time per HMC sample generation (Fig.~\ref{fig:speed}: cyan triangles).

Finally, we look at the time to generate independent samples in our chain and how it scales with increasing PTA size. This is ultimately the most important metric for testing the efficiency of HMC as independent samples and thinned Markov chains are what inevitably drive the statistical inferences made on the data. Independent samples in this context are defined here as samples that are separated by one autocorrelation length.

\begin{figure}
    \includegraphics[width=\columnwidth]{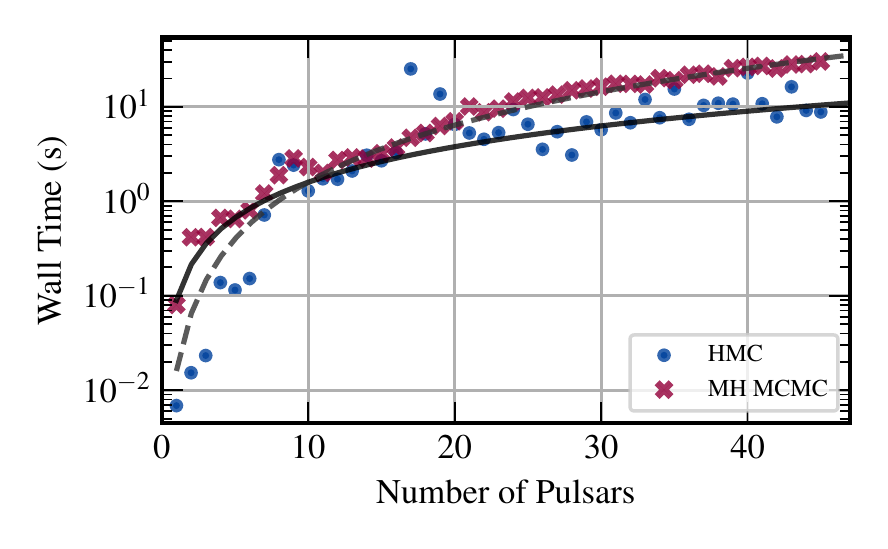}
    \caption{Wall time to produce an independent sample in Markov chains generated using HMC and MH MCMC methods, scaled by the number of pulsars $N_{\text{psr}}$ present in the model. The total number of parameters in a given model is $d = 2N_{\text{psr}} + 1$. The solid black represents the expected scaling for HMC of $\mathcal{O}(d^{5/4})$. The dashed gray line denotes the expected scaling for MH MCMC of $\mathcal{O}(d^{2})$.}
    \label{fig:speed_ac}
\end{figure}

We take the 45 PTA objects of increasing $N_{\text{psr}}$ and generate Markov chains with HMC and MH MCMC of size $M=8,000$ and $M=1,000,000$, respectively. Taking the median autocorrelation length of each chain and the base sampling speed calculated previously, we create a scaling of the wall time for both sampling methods in making independent samples for PTA models of increasing size. The results are summarized in Fig.~\ref{fig:speed_ac}. It shows that in the long run HMC will outperform MH MCMC techniques in making statistically relevant samples, despite the likely increase in upfront computational cost.

\section{Conclusions}
\label{sec:conclusions}

In this paper, we have implemented an efficient method for sampling the high dimensional distributions present in PTA GW Bayesian searches using the HMC algorithm. This method leverages a hybrid technique comprised of parts from both traditional stochastic Monte Carlo schemes as well as deterministic sampling methods derived from Hamiltonian dynamics. We show that utilizing HMC results in a reduction of approximately 2 orders of magnitude in the number of samples drawn to produce equivalent results to the existing Bayesian searches performed on PTA datasets.

The efficiency of this technique is largely defined by the speed at which derivatives of the log likelihood can be computed for the purpose of simulating Hamiltonian dynamics. We have shown that the current implementation of this calculation scales similarly to the present log-likelihood calculation with respect to the number of pulsars in a dataset, and improves upon traditional MCMC methods when comparing the production of independent samples of the distribution. This improvement in performance scaling is paramount because PTAs will continuously grow and add more pulsars to their data collection. The 11-year dataset featured in this paper contains 45 pulsars. Future NANOGrav datasets will have $\geq\!\!60$ pulsars and future IPTA datasets may contain close to $\sim\!\!100$ pulsars. Increasing the data volume will further strain our computational capabilities to perform large parameter GW searches. HMC provides a way of resolving these limitations in a way that is more favorable to future PTA analyses.

It is worth emphasizing that the $\mathcal{O}(d^{5/4})$ scaling of HMC is not just with respect to the number of pulsars, but with respect to the total number of parameters. We analyzed a model with $2N_{\mathrm{psr}} + 1$ parameters (with the GWB spectral index held fixed), but this represents only one of many different approaches for GW searches in PTAs. For example, one can parametrize the GWB with a free spectrum model, increasing its number of parameters from 2 to 30. Likewise, one can parametrize the individual pulsar red noise in a similar fashion, increasing the parameter count from 2 to 30 \textit{per pulsar}. The favorable scaling of HMC opens the door for more flexible models that are currently prohibitive with current MH MCMC runs.

Currently we have only applied the HMC algorithm to the problem of sampling a stochastic GWB model. PTAs are also sensitive to certain deterministic GW signals, and work towards tailoring this method to such searches is under development. This technique is particularly promising for searches for GWs from individual SMBHBs because of the large number of parameters necessary to describe the GW signal ($2 N_{\text{psr}} + 8$ for a circular binary, more if the source is eccentric). In general, this technique can be adapted to the full suite of PTA searches, provided the underlying models adhere to the limitations outlined in Sec.~\ref{sec:hmc}. The ultimate goal is a general purpose pipeline for performing any such PTA analysis that leverages the benefits of the HMC algorithm towards exploring complicated, high-dimensional models.

\acknowledgements
We thank Michele Vallisneri for useful discussions. We also thank Paul Baker for valuable comments. Lastly we thank the anonymous referees for their helpful comments and suggestions that improved the manuscript. This work was supported by National Science Foundation (NSF) Grant No. PHY-2011772. The authors are members of the North American Nanohertz Observatory of Gravitational Waves (NANOGrav) collaboration, which receives support from NSF Physics Frontiers Center Awards No. 1430284 and No. 2020265. G.E.F. is supported by National Aeronautics and Space Administration (NASA) Future Investigators in NASA Earth and Space Science and Technology Grant No. 80NSSC22K1591. A.D.J. and S.J.V. were supported by University of Wisconsin-Milwaukee (UWM) Discovery and Innovation Grant No. 101X410. A.D.J. acknowledges support from the Caltech and Jet Propulsion Laboratory President’s and Director’s Fund. This material is based upon work supported by NASA under Award No. RFP22\_5-0 issued through the Wisconsin Space Grant Consortium and the National Space Grant College and Fellowship Program. Any opinions, findings and conclusions or recommendations expressed in this material are those of the author and do not necessarily reflect the views of the National Aeronautics and Space Administration.

\bibliographystyle{apsrev}
\bibliography{authors}

\end{document}